# Verification of Turbulence Models for Flow in a Constricted Pipe at Low Reynolds Number

**Fardin Khalili***, Peshala P.T. Gamage, Hansen A. Mansy

[1]University of Central Florida, Department of Mechanical and Aerospace Engineering, Biomedical Acoustics Research Lab, Orlando, FL, 32816, USA

## ABSTRACT

Computational fluid dynamics (CFD) is a useful tool for prediction of turbulence in aerodynamic and biomedical applications. The choice of appropriate turbulence models is key to reaching accurate predictions. The present investigation concentrated on the comparison of different turbulence models for predicting the flow field downstream of a constricted pipe. This geometry is relevant to arterial stenosis in patients with vascular diseases. More specifically, the results of Unsteady Reynolds-Averaged Navier-Stokes (URANS) and scale resolving simulation (SRS) turbulence models such as Large Eddy Simulation (LES) and Detached Eddy Simulation (DES) were compared with experimental measurements. Comparisons included the mean flow and fluctuations downstream of the constriction. Results showed that the LES model was in better agreement with the velocity measurements performed using a Laser Doppler Anemometry (LDA). In addition, although URANS models predicted a wake region size and mean flow velocities comparable to SRS turbulence models, no small-scale vortical structures can be observed in the URANS solution due to the nature of these models. Modeling of these structures would, however, be helpful when more detailed flow behavior is needed such as in studies of acoustic sources. Hence, LES would be an optimal turbulence model for the flow under consideration, especially when sound generation would be of interest.

## 1. INTRODUCTION

One of the most important and least understood aspects of turbulence is when the local Reynolds number of the turbulence is relatively low [1]. Effects of turbulence is of interest in various industrial [2–6] and biomedical [7–11] applications. Most fluid flows are characterized by irregularly fluctuating flow quantities that often occur at small scales and high frequencies. Hence, resolving these fluctuation in time and space requires excessive computational costs. Optimum modeling of these structures is of interest for the acoustic investigations including biomedical applications, which are active areas of research [12–16]. Some basic knowledge of turbulence and an understanding of how turbulence models are developed can help provide insight into choosing and applying these models to obtain reasonable engineering simulations of turbulent flows.

Four turbulence models: two Reynolds-Averaged Navier-Stokes (RANS) models and two scale-resolving simulation (SRS) methods, namely, Large Eddy Simulation (LES) and Detached Eddy Simulation (DES), were included in the current study. RANS turbulence models solve for mean flow quantities where fluctuations are represented by ensemble averaging. On the other hand, LES simulates transitional flow with appropriate subgrid scale modelling, and was included here using Smagorinsky subgrid scale formulation. In addition, DES hybrid models incorporate LES modelling of free stream flow with unsteady RANS simulation of near wall flow, and are therefore less computationally expensive than LES. The main concern about the initial form of DES model was its inability to predict the behavior of the flow downstream of the separation region and improper simulation of laminar-turbulent transition [17,18]. In the last decade, significant developments in the DES modeling have resulted in improvements especially in solving the external flows in separation and strong circulation zones [19–21]. To highlight the main differences between these approaches, a brief descriptions of each turbulence model used in the current study is below.

*Corresponding Author: fardin@knights.ucf.edu

## 1.1 Reynolds-Averaged Navier-Stokes SST k-ω Model

The k- ω turbulence model is a two-equation model that solves transport equations for the turbulent kinetic energy ($k$) and the specific dissipation rate (ω), which is the turbulent dissipation rate (ε) per unit turbulent kinetic energy (ω ∝ ε/k). The k-ω two-equation model for low-Reynolds number flows was first proposed by Wilcox [22] and revised to better predict low-Reynolds number and transitional flows. The revised model accounted for several perceived deficiencies of the original version such as extreme sensitivity to inlet boundary conditions for internal flows [23]. The advantage of this model over the k-ε model is its improved performance for boundary layers under adverse pressure gradients [22,24]. On the other hand, the k-ε two-equation turbulence model, which solves transport equations for the turbulent kinetic energy and turbulent dissipation rate to calculate the turbulent viscosity, is more robust in wake regions and free shear flows [25]. These distinct capabilities led to the development of an integrated model that takes advantages of both models. That model was named shear stress transport (SST) k-ω model developed by Menter [26]. The SST k-ω model is essentially a k-ω model near wall boundaries and is equivalent to a transformed k-ε model in regions far from walls controlled by blending functions ($F_2$), (see, Eq. 4). In this model, the turbulent viscosity ($\mu_t$) is calculated as:

$$\mu_t = \rho k T \tag{1}$$

where, ρ is density, $k$ is turbulent kinetic energy, and T is the turbulent time scale. The turbulent kinetic energy can be defined as:

$$k = \frac{3}{2}(UI)^2 \tag{2}$$

where, U is the initial velocity magnitude, and I is initial turbulence intensity. In addition, the turbulent time scale in Eq. 1 can be calculated using Durbin's realizability constraint as:

$$T = min\left(\frac{\alpha^*}{\omega}, \frac{\alpha}{S F_2}\right) \tag{3}$$

where, S is the mean strain rate tensor. In this equation, $\alpha^*$ and $\alpha$ are model coefficients equal to 1 and 0.3, respectively [26]. $F_2$ can also be defined as:

$$F_2 = tanh((max\left(\frac{2\sqrt{k}}{\beta^*\omega d}, \frac{500v}{d^2\omega}\right))^2) \tag{4}$$

Where, $k, \omega, v$, and d are turbulent kinetic energy, specific dissipation rate, kinetic viscosity, and distance to the wall, respectively [26]. $\beta^*$ is the model coefficient:

$$\beta^* = F_1\beta_1^* + (1 - F_1)\beta_2^* \tag{5}$$

where, $\beta_1^*$ and $\beta_2^*$ are equal to 0.09, and $F_1$ can be illustrated as:

$$F_2 = tanh([min(max\left(\frac{\sqrt{k}}{0.09\omega d}, \frac{500v}{d^2\omega}\right), \frac{2\sqrt{k}}{d^2 CD_{k\omega}})]^4) \tag{6}$$

where, $CD_{k\omega} = max(\frac{1}{\omega}\nabla k.\nabla\omega, 10^{-20}$ is cross-diffusion coefficient.

## 1.2 Reynolds-Averaged Navier-Stokes (RANS) Reynolds Stress Transport Model

The development and application of Reynolds stress models can be traced back to the 1970s [27,28]. This model, also known as the second-moment closure model, directly calculates all components of the specific Reynolds stress tensor by solving governing transport equations, instead of calculating turbulence eddy viscosity. Hence, this model has the potential of predicting complex flows more accurately than two-equation models such as K-ε and K-ω turbulence models. This is due to the fact that the transport equations or the Reynolds stresses naturally account for the effects of turbulence anisotropy, streamline curvature, swirl rotation and high strain rates [29–31]. In this model, the turbulent viscosity is computed as:

$$\mu_t = \rho C_\mu \frac{k^2}{\varepsilon} \tag{7}$$

Where, $\rho$ is density, $\varepsilon$ is isotropic turbulent dissipation, and $C_\mu$ is the model coefficient equal to 0.09. The turbulent kinetic energy ($k$) can be defined as:

$$k = \frac{1}{2} tr(\boldsymbol{R}) \tag{8}$$

Where, $tr(\boldsymbol{R})$ represents the trace of Reynolds stress tensor (R). The tensor R can be written as:

$$R = \begin{bmatrix} \sigma_{xx} & \tau_{xy} & \tau_{xz} \\ \tau_{yx} & \sigma_{yy} & \tau_{yz} \\ \tau_{zx} & \tau_{zy} & \sigma_{zz} \end{bmatrix} = \rho \begin{bmatrix} \overline{u'u'} & \overline{u'v'} & \overline{u'w'} \\ \overline{v'u'} & \overline{v'v'} & \overline{v'w'} \\ \overline{w'u'} & \overline{w'v'} & \overline{w'w'} \end{bmatrix} \tag{9}$$

Where, $u'$, $v'$, and $w'$ are the velocity fluctuation components and, σ and τ represent normal and shear stresses, respectively.

## 1.3 Large Eddy Simulation (LES) Smagorinsky Subgrid Scale

This model is inherently a transient technique in which the large scales of the turbulence are resolved everywhere in the flow domain, and the small-scale motions are modeled (i.e., modeling less of the turbulence and explicitly solving for more of it to reduce the error in the turbulence modeling assumptions). LES is mainly used for flows with low Reynolds numbers [32,33]; therefore, to resolve the crucial turbulent structures near the wall, this approach requires an excessively fine mesh resolution [34]. This leads to high computational costs. The LES Smagorinsky Subgrid Scale provides the following mixing-length type formula for the subgrid scale viscosity [35]:

$$\mu_t = \rho \Delta^2 S \tag{10}$$

Where $\Delta$ is the length scale or grid filter width. The length scale $\Delta$ is directly related to the cell volume (V) and the wall distance, d, as follows:

$$\Delta = \begin{cases} f_v C_s V^{1/3} & \text{if length scale limit is not applied} \\ f_v \min\left(\kappa d, C_s V^{\frac{1}{3}}\right) & \text{if length scale limit is applied} \end{cases} \tag{11}$$

Where, $C_s$ is model coefficient of 0.1, and $\kappa = 0.41$ is von Karman constant [36]. In the above equations, $f_v$ is the Van Driest damping function. The turbulent eddy viscosity in standard Smagorinsky model is nonzero at solid boundaries and turbulence can be overestimated near the walls; hence the addition of this damping function handles this problem. More information about the applications of the Van Driest damping function in turbulence modeling can be found in [37].

## 1.4 Detached Eddy Simulation (DES) SST K-Omega

DES turbulence model is a hybrid approach where boundary layers and irrotational flow regions are solved using a RANS closure model while it will emulate a LES subgrid scale model in detached flow regions if the grid is fine enough [38]. In this way, the benefit of both models can be utilized: a RANS simulation in the boundary layers and an LES simulation in the unsteady separated regions [39,40]. In addition, this model incorporates the k-ω SST model as proposed by Menter [19] and is mostly appropriate for applications including complex recirculation systems and at high Reynolds numbers [41]. While DES holds great promise for certain types of simulations, it must be cautioned that it is not the answer to all turbulence modeling problems [42].

## 2. Models and Methods

### 2.1 Experimental Procedure

Laser Doppler Anemometry (LDA) measurements were performed at selected centerline locations (P1-P7 in Fig.1) downstream of the constriction to obtain a detailed representation of the flow characteristics. The entrance length of the pipe prior to the constriction was long enough to reach an approximately fully-developed turbulent air flow with the mean velocity of 0.89 m.s$^{-1}$ and the peak center-velocity of ~1 m.s$^{-1}$ at the inlet. The one-component velocity measurements were obtained with a LDA system (Dantec Dynamics A/S., Skovulunde, Denmark) used in backscatter mode. A Bragg cell was used to add an 80 MHz frequency shift to the beam with 660 nm wavelength. In addition, the two-component fiber optic transceiver (Model FlowExplorer; Dantec Dynamics A/S., Skovulunde, Denmark) with a 300-mm focal length lens was coupled to a fiber drive to produce an ellipsoidal probe volume with minor and major axes of 0.1 mm and 1 mm, respectively. High number of measurement samples (>~2000, with varying frequency between 200-1000 Hz) were acquired at each measurement location to ensure accurate results. The LDA system was calibrated for high-accuracy velocity measurements with calibration coefficient uncertainty lower than 0.1% (stated by Dantec Dynamics A/S., Skovulunde, Denmark) and mean confidence internal less than 0.09 m.s$^{-1}$.

### 2.2 CFD Modelling

The exact shape of an arterial stenosis varies from subject to subject and modeling it with accuracy is difficult. Hence a simplified stenosis shape will be considered in the current study. The schematic of the flow domain and the stenosis with the length of $L_c$ = 15 mm are shown in Fig. 1. An area reduction of 75% was chosen to model a moderate stenosis [43], where the pipe and stenosis inner diameters were D = 20.6 mm and $d_c$ = 10.3 mm, respectively. In addition, 7 equally-spaced (by 3 cm) points on the centerline of the pipe were chosen for velocity measurements.

The flow direction was set to z-direction in simulations as shown by an arrow in Fig. 1. In the current CFD simulations, the entrance length of the tube upstream of the constriction was $L_1$ = 20 mm with a constant mean inlet velocity ($\bar{U}_{inlet}$) of 0.89 ms$^{-1}$, which equals the mean inlet velocity measured with LDA. The outlet boundary condition was set to zero pressure ($P$ = 0). The density and dynamic viscosity of air were set to ρ = 1.184 kg.m$^{-3}$ and μ = 1.855E-5 Pa.s, respectively. These led to an inlet peak Reynolds number ($Re_{inlet} = (\rho \bar{U}_{inlet} D)/\mu$) of ~1170, and turbulent intensity of 5% (similar to the value measurement in the experiment). For the initial conditions, pressure was set to 0 Pa. Also, initial velocity was changed to a value close to inlet velocity to reduce the initial residual errors. The impact of the glass tube roughness was not considered in these simulations. Research on low Reynolds number flows is less commonly found, although they are highly relevant to biomedical applications. These flows with such low Reynolds number are considered turbulence as created through glottis in the upper airways [44]. The boundary conditions as well as turbulence parameters in this simple model are defined based on the previous studies in this field.

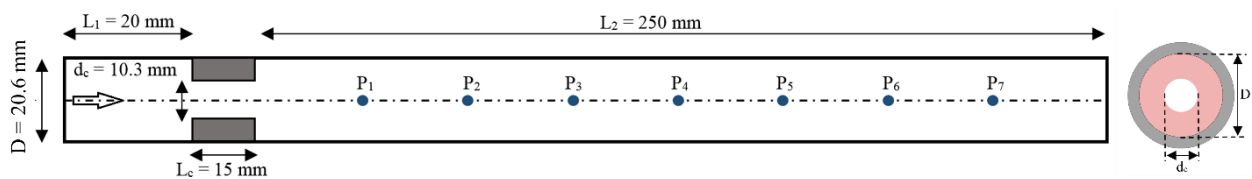

**Fig. 1** Side (left) and cross-sectional (right) views of the flow domain with 75% constriction.

This study was conducted using the commercial CFD code STAR-CCM+ (CD-adapco-Siemens, TX, USA) to compare the different turbulence methods. A time step of 0.0001 sec was used to ensure adequate time step convergence (less than $10^{-4}$ for all residuals), which was particularly important with SRS models [45]. All simulations used 2nd-order spatial and temporal discretization accuracy for all equations.

## 2.3 Mesh Configuration

The use of SRS turbulence models for wall bounded flows, requires high quality mesh. When creating such mesh, it is important that y+ ≤ 1 [8,46]. Polyhedral mesh was generated throughout the flow domain, with a refined mesh at the constriction and wake regions, Fig. 2a. This led to a mesh containing ~2 million hexahedral cells. In addition, accurate prediction of pressure drop in flows with separation depends on resolving the velocity gradients normal to the wall, as prism layers allow the solver to resolve near wall flow accurately [47,48]. Hence, a 5-layer prism layer mesh with a total thickness of 0.0003 m and layer stretching factor of 1.5 was employed near the boundaries, as shown in Fig. 1, to resolve the velocity gradients normal to the wall. The Y+ value was maintained in the order of 1 for all turbulence models chosen in this study. In addition, a mesh-independent study was conducted to find the optimized mesh configuration, Fig. 2b. Four different mesh configuration Mesh 1, 2, 3, and 4 were set up with approximately 700k, 1.4M, 2M, and 2.3M number of mesh cells. The evaluation of the mean velocity in flow direction (z-direction in the simulation) along the pipe showed that the Mesh 3 and 4 configurations led to similar results. Therefore, Mesh 3 was chosen as the optimized mesh.

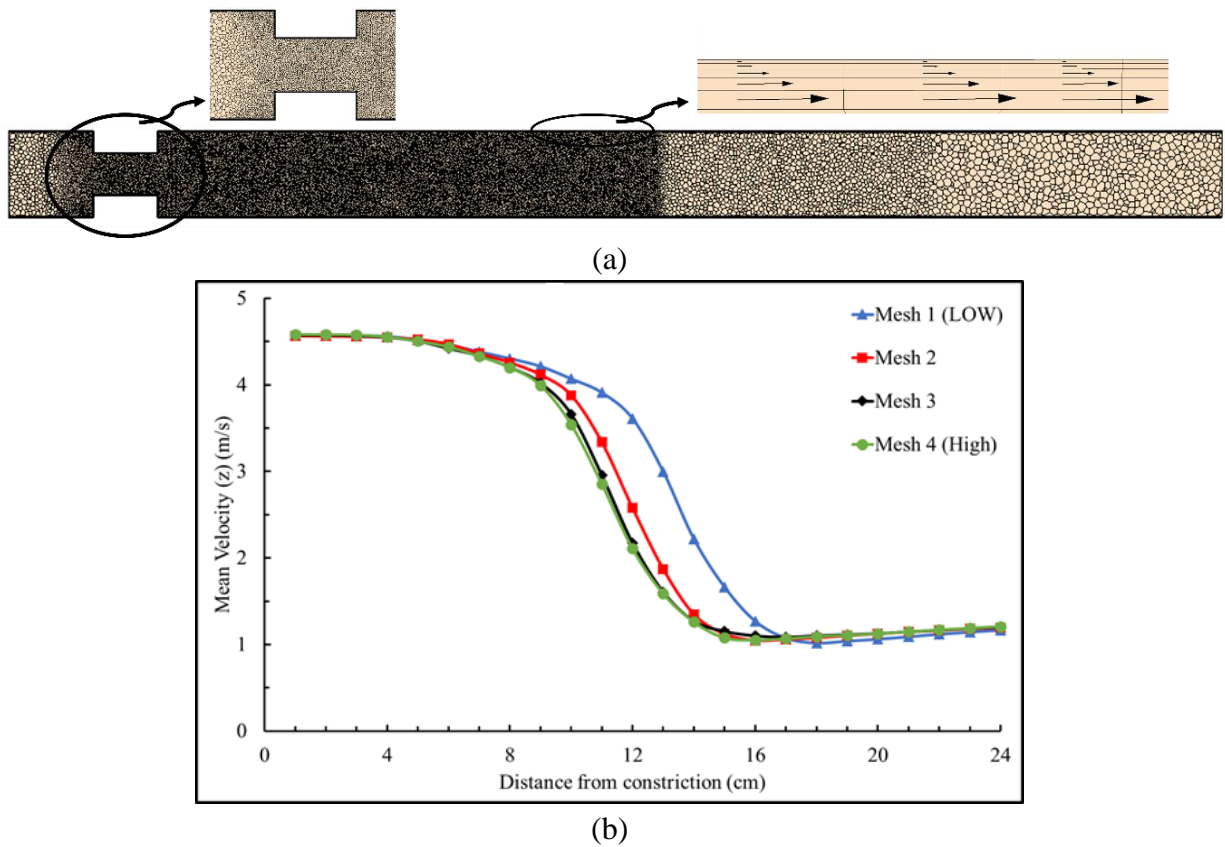

(a)

(b)

**Fig. 2** (a) Polyhedral mesh configuration with a refined mesh at and downstream of the constriction; (b) mesh-independent study for the current study

## 3. Results and Discussion

Fig. 3 shows the mean velocity measurements in the flow direction (z-component) at 7 different locations (i.e., P1-P7 in Fig 1) downstream of the constriction using LDA. The velocity measurements were performed for 30 seconds at each location. Three seconds of data were included in this figure to more clearly show the flow fluctuations. P1 and P2 were located in the "jet region" of the constriction and had the highest mean velocities and lowest fluctuations. The mean velocities at P1 and P2 were 4.66 and 4.63 m.s⁻¹ while the root-mean-square (RMS) of the velocity fluctuations were 0.047 and 0.115 m.s⁻¹, respectively. On the other hand, the highest fluctuations along with a significant drop in the mean velocities were observed at the next three centerpoints (P3, P4, and P5). Here, the mean velocities were 2.09, 1.17, and 1.11 m.s⁻¹, and RMS fluctuations were 0.708, 0.503, and 0.205 m.s⁻¹ at P3, P4, and P5, respectively. In the current study, this region with the highest fluctuations is called the "fluctuting zone". It can be noted that the RMS of the fluctuations decreased from P3 to P5 where the flow reattachment happened till it approaches fully developed/stable conditions at P6 and P7 in the "flow stabilization" region close to the outlet. The mean velocities at P6 and P7 were 1.14 and 1.16 m.s⁻¹ while the RMS of the fluctuations were 0.126 and 0.093 m.s⁻¹, respectively. Further measurements showed that the starting point of the high fluctuations levels was between P2 and P3. It was also observed that the "jet region" (characterised by high centerline velocity) extended up to about 7 cm downstream of the constriction while the "fluctuating zone" extended beyond this point. The highet value of the RMS velocity fluctuations (strongest turbulent stresses) was seen at 9 cm downstream of the constriction (P3).

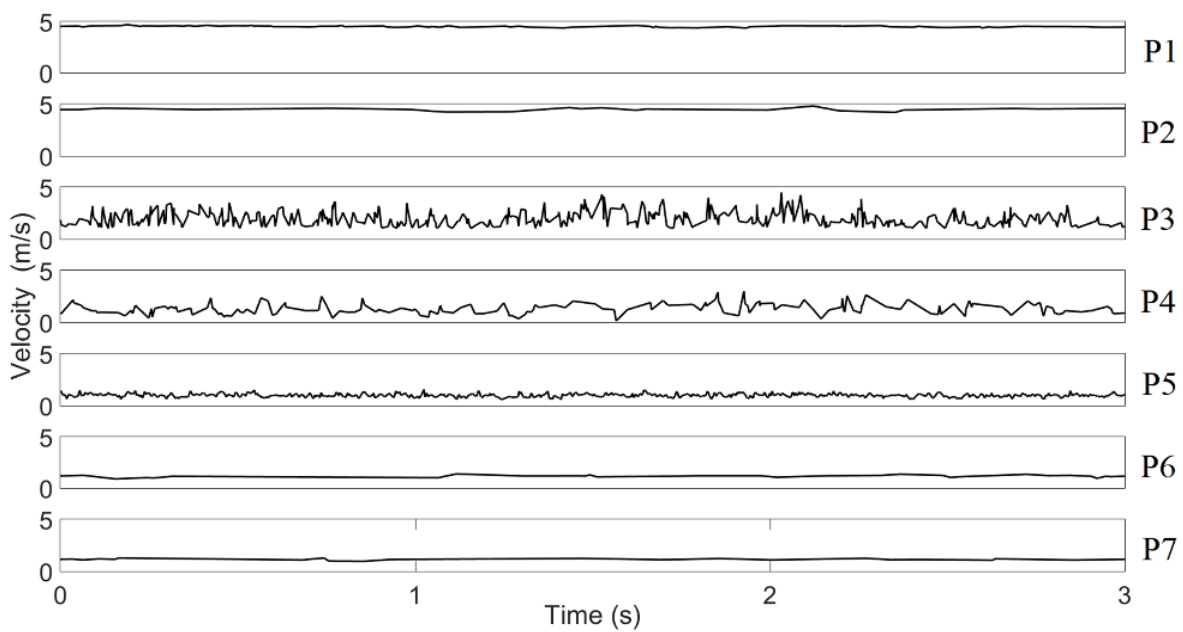

**Fig. 3** Centerline velocity measurements in the flow direction (z-components) using Laser Doppler Anemometry (LDA) at different locations (P1-P7) downstream of the constriction

Fig. 4a shows the z-component of mean velocity distribution for the different turbulence models while Fig. 4b shows the mean centerline velocities predicted by these models compared with the LDA measurements at P1-P7. In Fig 4a, the upper and lower regions of the flow jet with negative mean velocities represent the "recirculation zone". As the RANS turbulence models solve for mean flow quantities, they showed agreement with the mean velocity measurements. Agreement was highest at most measurement points (P1, P2, P6, and P7 in Fig. 4b), where the flow fluctuations were lowest. DES SST K-Omega results were also comparable. However, it can be seen in Fig. 4a that the jet high-speed zone extended to about P3 and P4 for RANS and P5 for DES. This is in contrast with LDA measurements that suggested that the jet high speed flow does not exist much beyone P2. Fig. 4b shows that the average velocity was least accurate for the DES SST K-Omega models in the "fluctuating zone" from P3 to P5. On the other hand, the LES model showed better agreement with the experiment than RANS models where K-ω was less accurate than the Reynolds stress model.

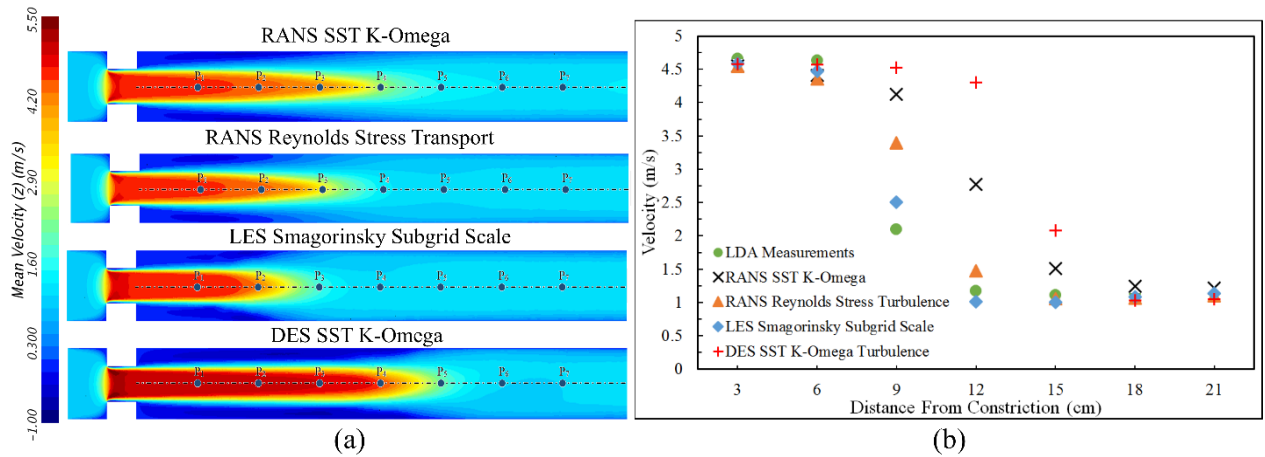

(a)                                                    (b)

**Fig. 4** Mean Velocity in the flow direction, a) on the side view of the flow domain for different turbulence models; b) on the centerline downstream of the constriction compared to the LDA measurements

The current study focussed on comparing the accuacy of different turbulence models for predicting flow characteristics in a duct with constriction. High CFD analysis precision is required for many applicatnins including flow induced sounds. Here, vortical structures containing significant fluctution levels are considered a main source of sound. Therefore, identificaion of optimal turbulence model that can predict flow fluctuations is important for such studies. Fig. 5 shows the RMS velocity fluctuations and voriticy for the four turbulence models under consideration. A vorticity range of 0.1 to 8000 s$^{-1}$ is displayed to more clearly show the detailed vortical structures in the "fluctuating zone". Results showed that RANS models do not resolve the turbulent fluctuations as suggested by previous studies [49]. In addition, the level of turbulent stresses and large-scale vortical structures in the detached shear layers emanating from flow separation are known to be underpredicted in RANS simulations [17,50], as can be seen in Fig. 5. Velocity fluctuations and vorticity were resolved better with DES than RANS models. DES did not estimate the right location of the "fluctuating zone" properly possibly due to the mesh configuration that was not fine enough for this model. This could have delayed the DES transition from RANS to LES far downstream of the separation line and underpredicted the the flow fluctuations.. Similar effects were discussed in more detail elsewhere [18]. On the other hand, LES provided the best results by showing the appearance of the RMS velocity fluctuations starting at 7 cm downstream of the constriction. In addition, LES predicted the occurrence of vortical structures including small and large eddies.

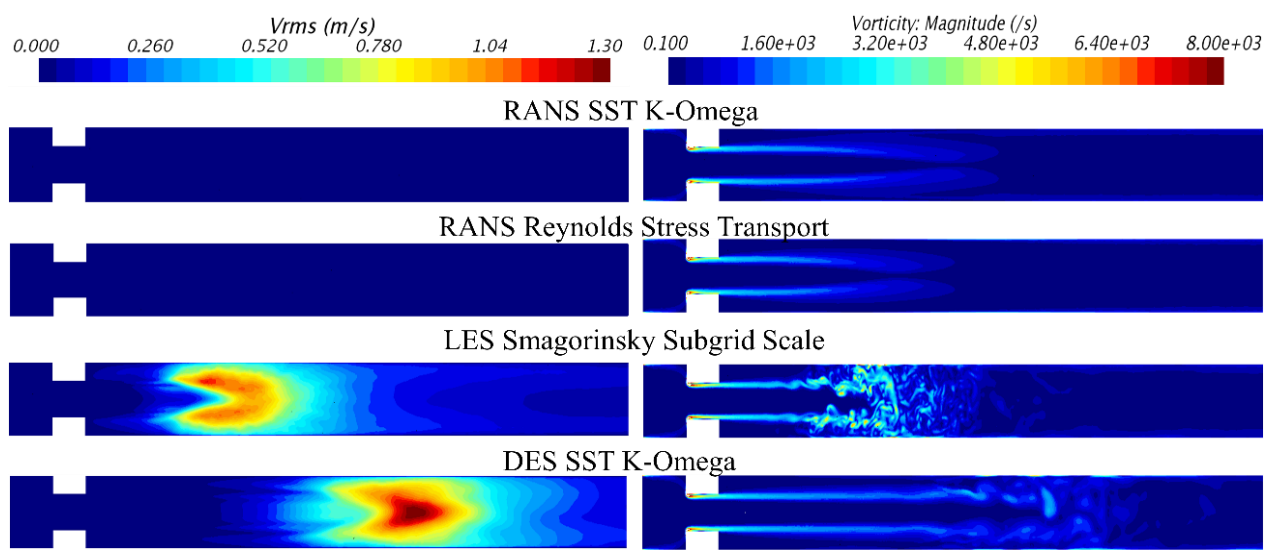

**Fig. 5** RMS of velocity fluctuations and vorticity at the flow domain cross-section

Fig. 6 displays the z-component of instantaneous velocity and streamlines using the LES turbulence model. The different flow regions can be seen including the jet and separation, fluctuating, flow reattachment, and flow stabilization regions, which are similarly labeled to an earlier study [51]. The velocity range in the figure is from 1.00 to the maximum instantaneous velocity of 5.41 m.s⁻¹ to help show the flow in the core of as well as the rest of the flow domain.

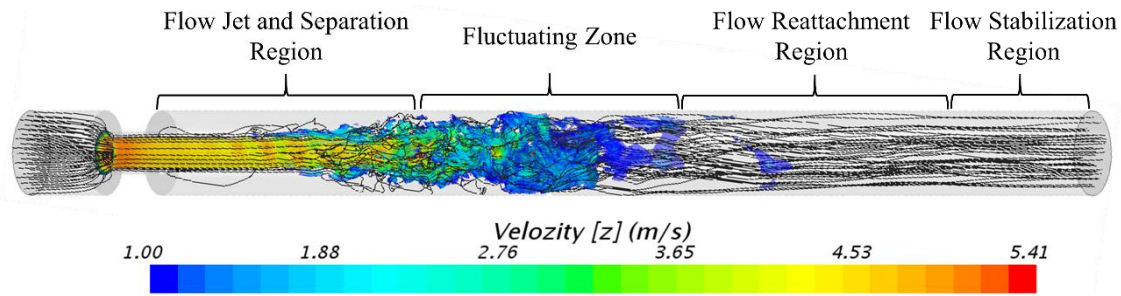

**Fig. 6** Discrete flow regions of flow jet and separation, fluctuating zone, flow reattachment, and flow stabilization regions which accurately captured by LES.

The order of the computational costs (the CPU time needed for the entire simulations) for such analysis were found to be for LES, DES, and RANS models. However, for such simple geometries the differences were not significant.

## 4. CONCLUSIONS

The current study assessed different turbulence models for the prediction of the flow field downstream of a constricted pipe. Models included: Unsteady Reynolds-Averaged Navier-Stokes (URANS) and scale resolving simulation (SRS) turbulence models such as Large Eddy Simulation (LES) and Detached Eddy Simulation (DES). Simulation results were compared with experimental measurements using a one-dimensional laser Doppler anemometry (LDA). Analyses of the mean velocity and local velocity fluctuations indicated that the LES Smagorinsky subgrid scale turbulence model had the highest agreement with experimental results. The accuracy of LES in predicting mean flow was followed by that of RANS Reynolds Stress, RANS K-Omega, and then DES. The RANS models do not, resolve turbulent flow fluctuations and eddies that would be a main source of sound generation. This would limit the utility of RANS in aeroacoustic studies. Also, DES did not localize the "fluctuating zone" properly and underpredicted the flow fluctuations in the separation zone for this low Reynolds number flow. Therefore, LES would be an optimal turbulence model for internal flow with constriction, especially when sound generation would be of interest.